\documentclass[namedreferences]{solarphysics}
\usepackage{verbatim}  
\usepackage[hyperref,optionalrh,solaromanenum]{spr-sola-addons} 
\usepackage{graphicx}                    
\usepackage{color}                       
\usepackage{breakurl}                         
\usepackage[flushleft]{threeparttable}
\usepackage{siunitx}
\usepackage{rotating}

\newcommand\apj{{Astrophys.~J.}}%
\newcommand\jgr{{J.~Geophys.~Res.}}%
\newcommand\solphys{{Solar~Phys.}}%
\newcommand\ssr{{Space~Sci.~Rev.}}%
\newcommand\apjl{{Astrophys.~J.~Lett.}}%
\newcommand\aap{{Astron.~Astrophys.}}%
\newcommand\apjs{{Astrophys.~J.~Supp.}}%
\begin{document}

\begin{article}

\begin{opening}

\title{How to Estimate the Far-Side Open Flux using STEREO Coronal Holes}

%
\author[addressref={mps},corref,email={stephan.heinemann@hmail.at}]{\inits{S.G.}\fnm{Stephan~G.~}\lnm{Heinemann}\orcid{0000-0002-2655-2108}}
\author[addressref={graz},corref,email={}]{\inits{M.}\fnm{Manuela~}\lnm{Temmer}\orcid{0000-0003-4867-7558}}
\author[addressref={col},corref,email={}]{\inits{S. J.}\fnm{Stefan~J.~}\lnm{Hofmeister}\orcid{0000-0001-7662-1960}}
\author[addressref={graz},corref,email={}]{\inits{J.}\fnm{Aleksandar ~}\lnm{Stojakovic}\orcid{}}
\author[addressref={mps,unigott,UAE},corref,email={}]{\inits{J.}\fnm{Laurent~}\lnm{Gizon}\orcid{0000-0001-7696-8665}}
\author[addressref={mps},corref,email={}]{\inits{J.}\fnm{Dan~}\lnm{Yang}\orcid{0000-0001-7570-1299}}
%
\runningauthor{S.G. Heinemann et al.}
\runningtitle{Far-Side Open Flux}

\address[id={mps}]{Max-Planck-Institut f\"ur Sonnensystemforschung, Justus-von-Liebig-Weg 3, 37077 Göttingen, Germany}
\address[id={graz}]{University of Graz, Institute of Physics, Universit{\"a}tsplatz 5, 8010 Graz, Austria }
\address[id={col}]{Columbia Astrophysics Laboratory, Columbia University, New York, USA}
\address[id={unigott}]{ Institut f\"ur Astrophysik, Georg-August-Universit\"at G\"ottingen,  37077~G\"ottingen, Germany}
\address[id={UAE}]{Center for Space Science, NYUAD Institute, New York University Abu Dhabi, Abu Dhabi, UAE}
\begin{abstract}
Global magnetic field models use as input synoptic data, which usually show ``aging effects" as the longitudinal $360^{\circ}$ information is not obtained simultaneously. Especially during times of increased solar activity, the evolution of the magnetic field may yield large uncertainties. A significant source of uncertainty is the Sun’s magnetic field on the side of the Sun that is not visible to the observer. Various methods have been used to complete the picture: synoptic charts, flux-transport models, and far side helioseismology. 
In this study, we present a new method to estimate the far-side open flux within coronal holes using STEREO EUV observations. First, we correlate the structure of the photospheric magnetic field as observed with the \textit{Helioseismic and Magnetic Imager} on board the \textit{Solar Dynamics Observatory} (HMI/SDO) with features in the transition region. From the $304$\AA\ intensity distribution, which we found to be specific to coronal holes, we derive an empirical estimate for the open flux. Then we use a large sample of $313$ SDO coronal hole observations to verify this relation. Finally, we perform a cross-instrument calibration from SDO to STEREO data to enable the estimation of the open flux at solar longitudes not visible from Earth. We find that the properties of strong, unipolar magnetic elements in the photosphere, which determine the coronal hole's open flux, can be approximated by open fields in the transition region. We find that structures below a threshold of $78\%$ (STEREO) or $94\%$ (SDO) of the solar disk median intensity as seen in $304$\AA\ filtergrams are reasonably well correlated with the mean magnetic flux density of coronal holes (cc$_{\mathrm{sp}} = 0.59$). Using the area covered by these structures ($A_{\textsc{of}}$) and the area of the coronal hole ($A_{\textsc{ch}}$), we model the open magnetic flux of a coronal hole as $|\Phi_{\textsc{ch}}| = 0.25 A_{\textsc{ch}} ~\mathrm{exp}(0.032 A_{\textsc{of}})$ with an estimated uncertainty of $40$ to $60\%$.

\end{abstract}

%

\end{opening}



\section{Introduction} \label{s:intro}
The complex solar magnetic field is usually thought of as an interplay of open and closed fields from the smallest spatial scales up to global scales. The portion of the magnetic field that affects interplanetary space is called open \textit{i.e.,} the fields close far away from the Sun. Most open fields, and the resulting open flux, is supposed to originate in large-scale magnetic structures known as coronal holes. These coronal holes are usually observed in the solar corona as regions of reduced emission in high-temperature wavelengths such as extreme ultraviolet (EUV) or X-ray \citep[see Review by][and references therein]{cranmer2009}. Along the open field, plasma is accelerated to speeds up to $\approx 800$\,km\,s$^{-1}$. These plasma outflows are known as high--speed solar wind streams and are the leading cause of minor to moderate geomagnetic activity at distances of 1AU \citep{wilcox68,1997farrugia,alves06, 2017vrshnak, 2018richardson}.\\ 

Intuitively, one would expect that the gross of the open magnetic flux from the Sun originates in coronal holes, however, comparisons of \textit{in-situ} measurements at 1AU and remote sensing observation of the Sun have revealed a significant discrepancy in the open flux of more than a factor of two, often referred to as the ``Open Flux Problem" \citep[][]{Linker_2017, 2017Lowder, 2019wallace}. There are several reasons for this: underestimation of open flux from individual coronal holes due to the method and data used \citep[][]{2021Linker_ISSI}; the limitation that only  near-side magnetic field observations are available;insufficient knowledge of the magnetic flux originating from the solar poles, \textit{i.e.,} polar coronal holes; and open flux possibly not associated with observed coronal holes. Recently launched and future space missions going out of the ecliptic plane, such as Solar Orbiter \citep{2020muller_solO} or SOLARIS \citep{2020hassler_solaris} will shed with their unprecedented observational data more light on the polar regions, particularly their magnetic fields. Nowadays, the far-side magnetic flux is estimated using synoptic charts \citep[\textit{e.g.,}][]{2017Lowder}, that only show the magnetic field over multiple days and weeks, flux transport models \citep[\textit{e.g.,}][]{2010Arge,2015hickmann}, helioseismic measurements adapted into flux transport modeling \citep[][]{2013Arge} or by estimating far-side active regions \citep[][]{2014Liewer}. Newer approaches try to use machine learning, utilizing helioseismology or are planning to use data from Solar Orbiters Polarimetric and Helioseismic Imager \citep[SO/PHI:][]{2020solanki_so_phi}, however, these methods are still in their infancies. \\

Coronal holes extend from the solar surface up into the corona and interplanetary space, and its magnetic field tracers may be seen in heights above the photosphere. On a global perspective, coronal holes appear as large scale structures that are dominated by one polarity which they retain over their lifetime \citep[][]{1982levine,1996wang, 2018heinemann_paperII,2020Heinemann_chevo}. Their open flux (signed to unsigned flux) varies between $6$ and $87\%$ \citep[][]{2017hofmeister,2019heinemann_catch}, the average flux densities are $1-15$ Gauss ($\approx3$G: \citealt{Bohlin1978}, $1-7$G: \citealt{1982Harvey}, $1-5$G: \citealt{1989Obridko}, $3.0\pm1.6$G: \citealt{2017hofmeister} and $2.9\pm1.9$G: \citealt{2019heinemann_catch}), and they have a significantly reduced temperature ($\approx 0.95$MK) and density ($\approx 1.5-2.5 \times 10^{8}$ cm$^{-3}$) in contrast to the surrounding quiet Sun \citep[][]{1999warren,2011hahn,2018Wendeln, 2020saqri, 2021heinemann_dem}.  The magnetic fine structure of coronal holes is defined by a very weak background field associated with the insides of supergranular cells and unipolar centers of flux accumulation occurring at the lanes and nodes of the magnetic network (\textit{i.e.,} magnetic elements). These magnetic elements may form open funnels that guide the open magnetic flux and solar wind, or form closed loops with nearby magnetic elements of opposite polarity \citep[][]{1973dunn, 2001berger, cranmer2009, 2019hofmeister}. \cite{2004wiegelmann} showed that a relatively large number of these closed structures exists in coronal holes, however, their average height is lower than in the quiet Sun, most likely due to the canopy-like structure of the opening funnels that inhibits the expansion of the closed loops in height. Therefore, the fine scale structure of magnetic field within coronal holes leaves traces in larger heights, which possibly can be identified in observations of the transition region such as He \textsc{ii} $304$\AA\ EUV data \citep[\textit{e.g.,}][]{2013Jin}. The $304$\AA\ line is dominated by He \textsc{ii} emission at a peak response temperature of about $50$kK which images primarily the upper chromosphere and transition region \citep[][]{2012lemen_AIA}. This is also the region where it is believed that the rapid expansion of the coronal funnels takes place, and this makes it possible to reliably study the open as well as the closed fields. Because of this and because the filter is available on both spacecraft used in this study, the $304$\AA\ line is suitable to study the relation between the photospheric magnetic field and the chromosphere/transition region intensity of coronal holes, and thereby to estimate the magnetic flux originating from far-side coronal holes. \\

In this paper, we propose a concept on how to estimate the far-side flux of the Sun using EUV observations of coronal holes taken by the SECCHI suite \citep[\textit{Sun Earth Connection Coronal and Heliospheric Investigation};][]{2008howard_SECCHI} on-board the STEREO \citep[\textit{Solar TErrestrial RElations Observatories;}][]{2008kaiser_STEREO} spacecraft. First, we analyze coronal hole properties in the photosphere and the transition region, and show how the open flux of coronal holes can be estimated in higher layers of the solar atmosphere. Second, we use this relation to derive an estimate of the open flux. For this we use near-side observations from the \textit{Helioseismic and Magnetic Imager} \citep[HMI;][]{2012schou_HMI,2016couvidat_HMI} and the \textit{Atmospheric Imaging Assembly} \citep[AIA;][]{2012lemen_AIA} on-board the \textit{Solar Dynamics Observatory} \citep[SDO;][]{2012pesnell_SDO}. After establishing the correlation, we use STEREO coronal holes observed at a relatively small separation angle to Earth to calibrate the relation obtained from SDO data for application on STEREO data. Last, we discuss the physical origin of this relation, the implication for space weather forecast and the limiting factors of the presented method.

\section{Methodology}\label{s:methods}

\subsection{Magnetic Flux and EUV}
\subsubsection{Coronal Holes}\label{subs:met-ch}
First, we relate near-side features of the photospheric magnetic field to the EUV intensity distribution of coronal holes. To derive statistically significant results, we use the coronal holes from the CATCH catalog\footnote{see \url{https://vizier.u-strasbg.fr/viz-bin/VizieR-3?-source=J/other/SoPh/294.144/}.} that contains $707$ well-defined non-polar coronal holes extracted from SDO/AIA $193$\,\AA\ filtergrams using \textit{Collection of Analysis Tools for Coronal Holes} (CATCH). The catalog includes coronal hole properties \textit{e.g.,} area, intensity, signed and unsigned magnetic field strength and magnetic flux including uncertainty estimates, and the extracted boundaries. The magnetic properties were derived from 720s Line-of-Sight (LoS) HMI/SDO magnetograms. To correlate properties of the photosphere to the transition region, we constrain our sample to coronal holes with an area of $A_{\textsc{ch}} \ge 2 \times 10^{10}$ km$^{2}$, \textit{i.e.,} all coronal holes that can be considered medium to large. This yields a statistically significant dataset of $313$ coronal holes, distributed between latitudes of $\pm 60^{\circ}$, located close to the central meridian and covering the timerange from 2010 to 2019. A detailed description of the catalog and the tool is given in \cite{2019heinemann_catch}. \\

We use these $313$ coronal holes to derive the intensity distribution in AIA/SDO $304$\AA\ filtergrams. The $304$\AA\ data was prepped to level $1.5$ (using the SSWIDL routine \texttt{aia\_prep}) as well as corrected for filter degradation using the routines provided by the instrument team. Additionally, due to the reduced emission from coronal holes as well as the generally low counts caused by the strongly degraded filter, for every coronal hole 30 images at a cadence of 12s ($=6$min) are stacked to increase the signal-to-noise ratio. Further, we correct for limb-brightening using an annulus limb-brightening correction \citep[][]{2014spoca}. Finally the AIA - $304$\AA\ data was co-registered to the coronal hole boundary maps and the regions of interest were extracted.

\subsubsection{Quiet Sun}\label{subs:met-qs}
To investigate the reliability and significance of the correlations found in coronal holes, we use a reference dataset consisting of $657$ subfields (sf) of a size of $150"\times150"$ in quiet Sun regions, defined as areas without active regions, coronal holes, filaments, and filament channels. The quiet Sun regions were selected to be spread over a wide latitudinal range similar to the coronal hole dataset, \textit{i.e.} $\pm 60^{\circ}$, to cover the time period from 2010 to 2019. The $304$\AA\ data was pre-processed analog to the coronal hole dataset as described in Section~\ref{subs:met-ch}. The average magnetic flux density of these regions was derived from 720s LoS HMI/SDO magnetograms following \cite{2017hofmeister} and \cite{2019heinemann_catch} \textit{e.g.,} correcting for the LoS observations under the assumption of radial field. We further divide the sample into quiet regions (defined as $|B_{\mathrm{sf}}| \leq 3$G) and plage regions (defined as $|B_{\mathrm{sf}}| > 3$G). This reference dataset is used to evaluate whether the correlation found for coronal holes is a general property of the transition region or specific to coronal holes.

\subsection{Cross-correlation to STEREO}
To determine the open flux of coronal holes on the solar far-side, we have to relate the AIA/SDO observations to observations taken by EUVI/STEREO. To do so, we use $13$ coronal holes that are visible in SDO, STEREO-A, and STEREO-B, within a few days of each other. The coronal holes were selected between October 2010 and July 2011, when the separation angle between Earth and STEREO was lower than $90^{\circ}$ in longitude and SDO was already launched. This spatially close position enables a calibration from SDO to STEREO, as in a first order approximation the evolution of coronal holes in this time range, especially during low solar activity, can be neglected \citep[][]{temmer18}. \cite{2020Heinemann_chevo} showed that the signed magnetic flux density changes on average at a rate of $27.3 \pm 32.2$ mG day$^{-1}$. From this we derive that the difference in magnetic flux density between SDO, where the magnetic field measurements were taken, and the STEREO observations of the coronal holes is $<0.5$G.\\

These 13 coronal holes are only partly covered by the CATCH catalog for the SDO observations and not covered for STEREO observations. Therefore, we detect and extract the coronal holes using CATCH \citep[][]{2019heinemann_catch} on 193\AA\ (AIA/SDO) and 195\AA\ (EUVI/STEREO) filtergrams. The magnetic field information was again derived from 720s LoS HMI/SDO magnetograms as described in Section~\ref{subs:met-qs}. The EUV intensity distributions were derived in analogy to Section~\ref{subs:met-ch} from level 1.5 $304$\AA\ EUVI/STEREO data (using \texttt{secchi\_prep}) and pre-processed 1.5 level $304$\AA\ AIA/SDO data.

\subsection{Error Estimation}\label{subs:corr_methods}
The analysis of different parameters \textit{e.g.,} \textit{Spearman} correlation coefficients and fits, was done using a bootstrapping method \citep{efron1979_bootstrap,efron93_bootstrap} to derive errors that take into account the sample size, outliers and uncertainties.

\section{Results}\label{s:res}
We investigated the relation between the magnetic flux density of coronal holes and the intensity distribution associated to open and closed fields around the transition region using $304$\AA\ data and obtained the following results.

\subsection{Photospheric Magnetic Field and Magnetic Elements}
\begin{figure} 
 \centerline{\includegraphics[width=.9\textwidth,clip=]{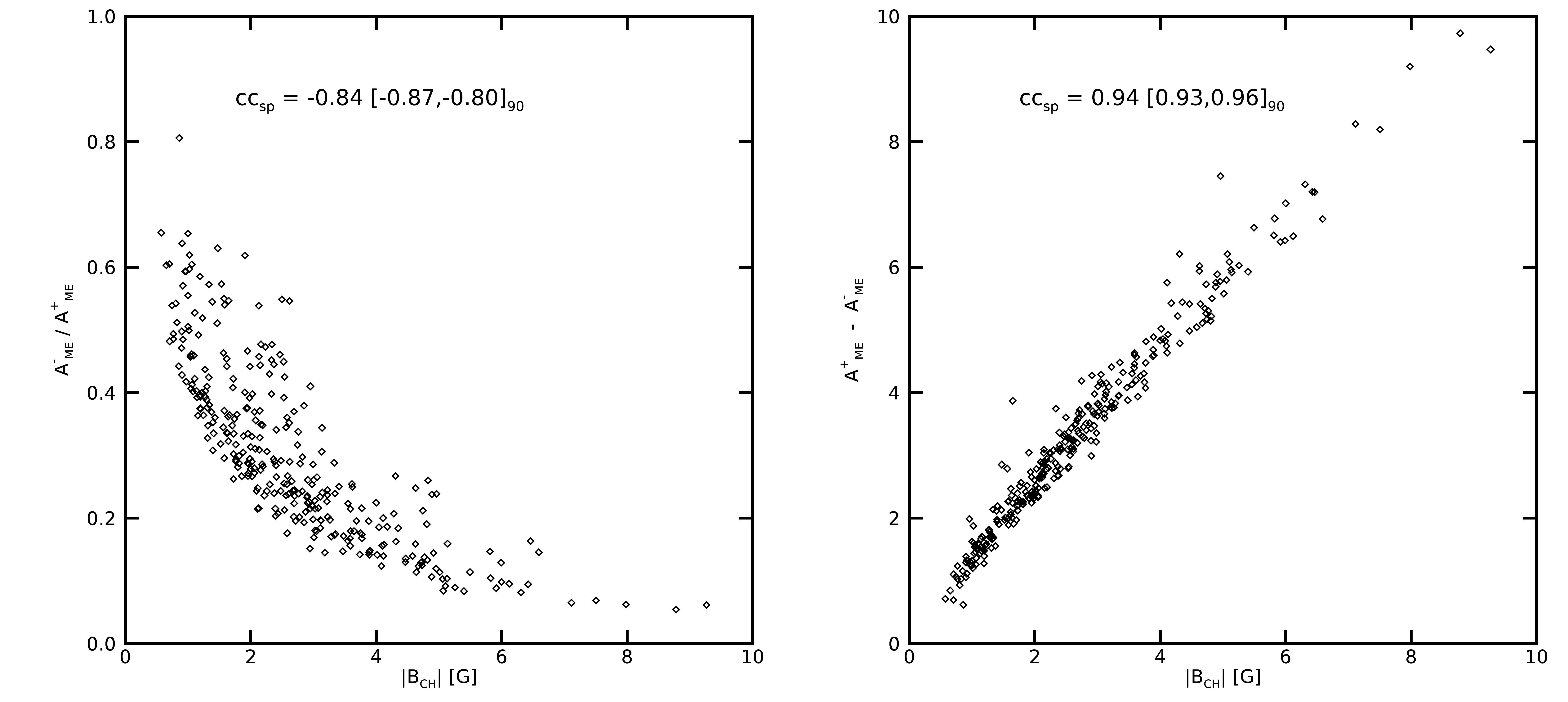}}
 \caption{Left: Ratio of the area of magnetic element of non-dominant polarity to area of magnetic elements of dominant polarity in coronal holes as function of the magnetic flux density. Right: Percentage area coverage of open magnetic elements as function of the magnetic flux density.}\label{fig:me_vs_b}
 \end{figure}

It has been shown by \cite{2017hofmeister} and further confirmed by \cite{2019hofmeister} and \cite{2019heinemann_catch} that the signed mean magnetic field strength (consequently open flux) of coronal holes is predominantly rooted in strong, long-living magnetic elements in the photosphere, that cover roughly $1$ to $10\%$ of the coronal hole area. Magnetic elements are either associated with the dominant polarity or the non-dominant polarity of the coronal hole. Following the work by \cite{2019hofmeister}, we first investigate the area they cover as function of the coronal holes mean magnetic flux density. In Figure~\ref{fig:me_vs_b}, we show the proportion of the coverage of non-dominant ($A^{-}$) to dominant polarity ($A^{+}$) magnetic elements as function of magnetic field strength. We find a strong \textit{Spearman} correlation, cc$_{\mathrm{sp}} = -0.84$ with a  $90\%$ confidence interval of CI$ = [-0.87,-0.80]$, that shows a coverage decrease towards stronger fields (left panel). The right panel shows the area that is covered by supposedly open fields as function of the magnetic flux density. We define open field as $A_{\textsc{of}} = A^{+}-A^{-}$ under the assumption that all non-polarity magnetic elements close with dominant polarity magnetic elements within the coronal hole. We find a very strong, close to 1-to-1 correlation, cc$_{\mathrm{sp}} = -0.94$, CI$_{90} = [0.93,0.96]$ between the percentage covered by open fields and the mean magnetic flux density of the coronal hole. The open field of coronal holes,  estimated by the area that magnetic elements of dominant polarity cover, builds the basis for estimating the magnetic flux in coronal holes from non-magnetic field data.

\subsection{Coronal Holes and the Transition Region} \label{subs:ch+tr}
\begin{figure} 
 \centerline{\includegraphics[width=.9\textwidth,clip=]{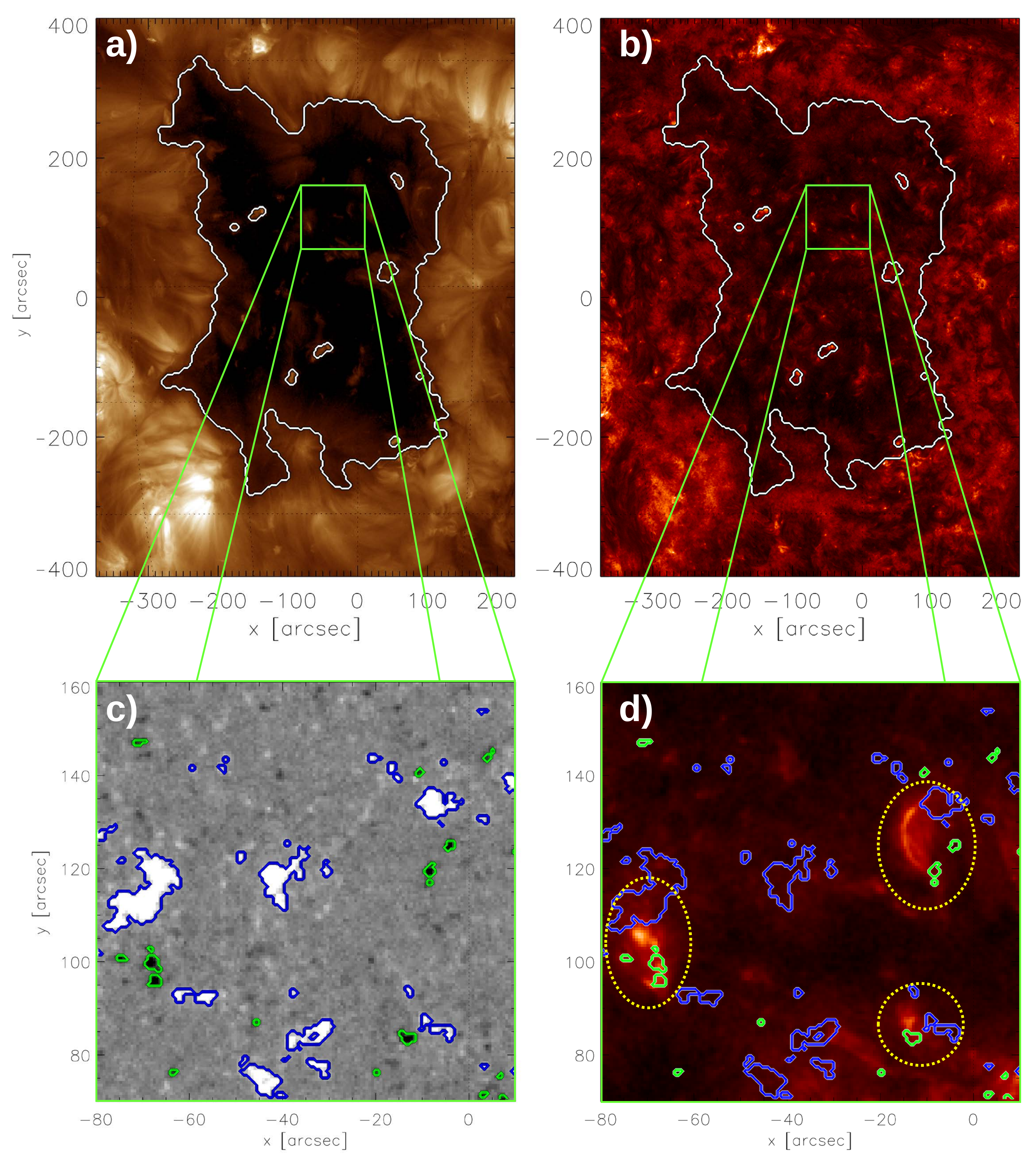}}
 \caption{Example of a coronal hole from May 29, 2013 12:00UT. a) shows the coronal hole in $193$\AA; b) show the $304$\AA\ filter, the white contour represents the coronal hole boundary as extracted by CATCH. c) shows the underlying photospheric magnetic field of a subfield as marked by the green line; d) shows the same subfield for the $304$ line. Magnetic elements of dominant polarity (in this coronal hole positive) are overlayed in blue and opposite polarity magnetic elements in green. The yellow circle mark bright structures in $304$\AA\ that can be associated to a bipolar set of magnetic elements in the underlying photosphere. Single magnetic elements of dominant polarity are not associated with bright structures.
 }\label{fig:pretty_img}
 \end{figure}

The magnetic field in coronal holes follows a particular, however not well known, expansion profile, where strong unipolar funnels rapidly expand into the open space before forming the approximately homogeneous vertical fields in the upper corona \citep{2005cranmer}. Closed fields of quiet coronal hole regions and small bipolar structures seem to be prevented from expanding to great heights \citep[][]{2004wiegelmann}. To relate the magnetic field in the photosphere to the field in the transition region, we assume that in the transition region, where the $304$\AA\ He \textsc{ii} line is primarily formed, the space is filled with either open fields or closed fields, both anchoring in the coronal holes magnetic elements. Thereby, the closed fields are most likely associated with loops of bipolar structures \textit{e.g.,} magnetic elements of opposite polarity and weaker and smaller loops are supposed to close at lower heights. Figure~\ref{fig:pretty_img} shows a coronal hole from May 29, 2013 12UT in the $193$\AA\ as well as the $304$\AA\ line. Within the small subfield we show the LoS magnetic field whereby we mark the dominant and non-dominant magnetic elements in blue and green. The elements are overlaid onto the $304$\AA\ subfield image, and it is clearly apparent that the bright structures in the transition region can be associated to bipolar footpoint pairs, in contrast, the large magnetic elements of dominant polarity are not associated with any specific structure. These images are representative for the entire dataset. \\

\begin{figure} 
 \centerline{\includegraphics[width=.9\textwidth,clip=]{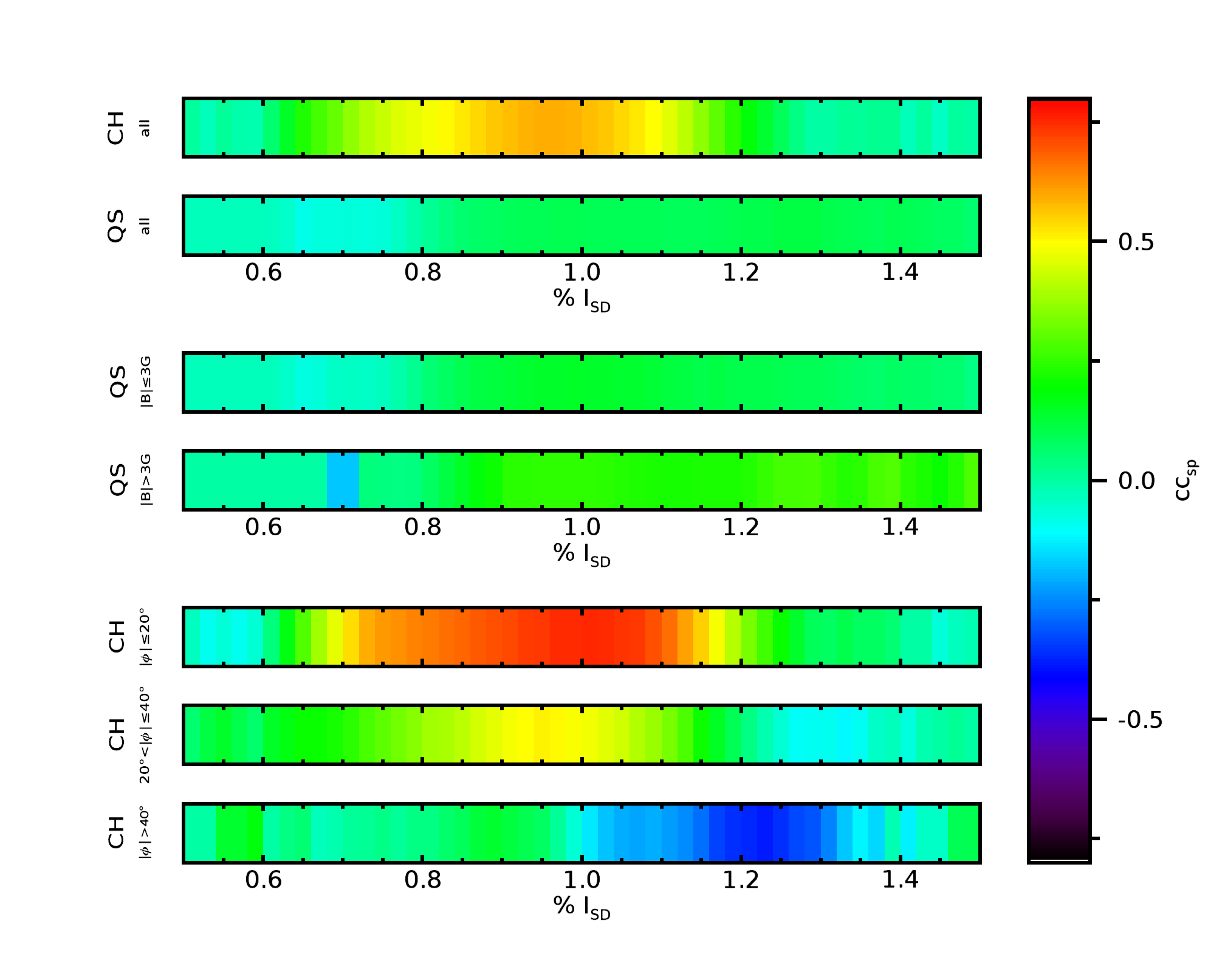}}
 \caption{Correlation of mean magnetic flux density ($|B_{\textsc{ch}}|$) of the coronal hole and the percentage area coverage ($A_{\textsc{of}}$) of the coronal holes (below a set intensity threshold) as function of the threshold normalized to the solar disk median intensity ($I_{\mathrm{SD}}$). The bars from top to bottom are sorted according to the data (sub-)set used. The top bar includes all coronal holes followed by all quiet Sun areas; The next two panels show the ``quiet" ($|B_{\mathrm{sf}}| \leq 3$) and the plage ($|B_{\mathrm{sf}}| > 3$) regions of the quiet Sun dataset; The bottom three panels show different coronal hole subsets sorted into latitude bins depending on the latitudinal position of their geometric center of mass.}\label{fig:corrbar}
 \end{figure}

Under the previously stated assumption that the space in the transition region is filled by either open, expanded fields from dominant polarity magnetic elements associated with lower $304$\AA\ intensities or closed fields from bipolar structures that can be clearly associated to bright structures, we segment the coronal hole area as seen in $304$\AA\ filtergrams using an intensity threshold into higher and lower intensity regions. As such we calculate the percentage area coverage of supposedly open fields, $A_{\textsc{of}} = \frac{A_{\textsc{thr}}}{A_{\textsc{ch}}}$ for a range of thresholds, whereby we define the threshold in percent of the solar disk median intensity ($I_{\mathrm{\textsc{SD},med}}$) to remove solar cycle effects, and correlate $A_{\textsc{of}}$ to the signed mean magnetic flux density of the coronal hole. In addition, this analysis was also performed on quiet sun and plage subfields to determine whether these are properties unique to coronal holes. This is shown in Figure~\ref{fig:corrbar}. The x-axis represents the upper threshold \textit{i.e.} the area below the threshold is considered ``open" fields, and the color shows the \textit{Spearman} correlation coefficient of  $A_{\textsc{of}}$ to $|B_{\textsc{ch}}|$ for coronal holes or  $A_{\textsc{of}}$ to $|B_{\mathrm{sf}}|$ for quiet Sun subfields. For coronal holes, we find a maximum in the correlation with cc$_{\mathrm{sp}} = 0.59$, CI$_{90} = [0.52,0.66]$  at a threshold of $94\%$ of the median solar disk (SD) intensity, whereas in quiet Sun regions, neither for actual quiet regions nor for plage regions, a clear correlation exists (nor any changes). The correlation for QS subfields does not exceed cc$_{\mathrm{sp}}= \pm 0.25$. This suggests that the previously stated assumption of open fields is valid for coronal holes but not quiet Sun regions, which was expected. Next, we sort the coronal holes into three bins depending on their latitudinal position of their geometric center of mass (CoM;  see Fig.~\ref{fig:corrbar}, bottom). From that we find a further latitudinal dependence. The strongest correlation was derived for latitudes $ < 20^{\circ}$ which already drops in the $ 20^{\circ} - 40^{\circ}$ bin and even ceases for latitudes $ > 40^{\circ}$. Possible reasons are discussed in Section~\ref{s:disc}.\\

\begin{figure} 
 \centerline{\includegraphics[width=.9\textwidth,clip=]{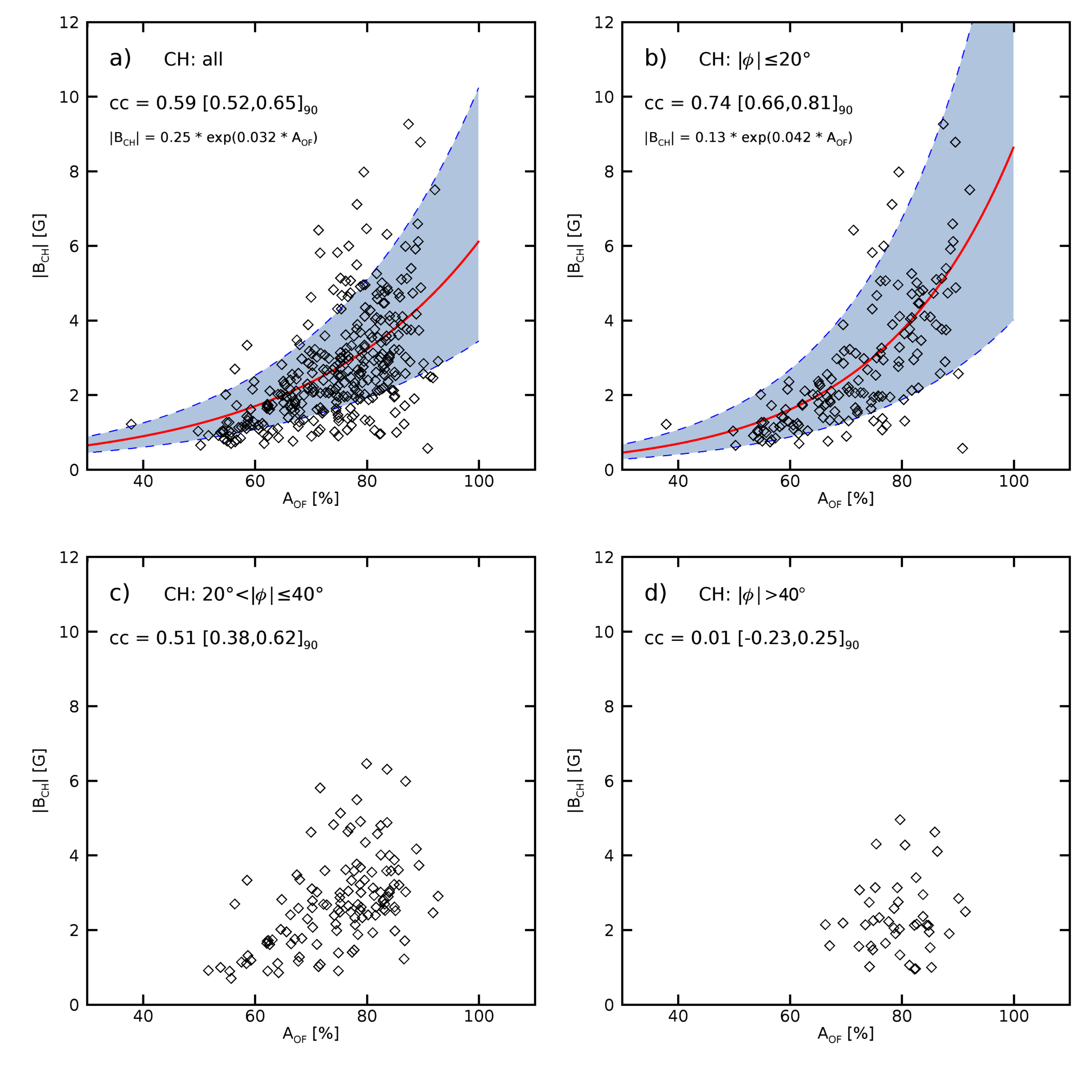}}
 \caption{Coronal hole mean magnetic flux density ($|B_{\textsc{ch}}|$)  as function of the percentage area coverage ($A_{\textsc{of}}$) of assumed open fields extracted using an upper threshold of $0.94 \times I_{\mathrm{\textsc{sd},med}}$. Panels a) through d) show all coronal holes as well as three subsets sorted by their latitudinal position. The red lines shows an exponential fit to the data and the shaded areas give the fit uncertainties.}\label{fig:corrplot}
 \end{figure}

By choosing the threshold associated with the best correlation (Figure~\ref{fig:corrbar}, top panel), we can analyze the relation in more detail. Figure~\ref{fig:corrplot} shows the absolute value of the signed mean magnetic flux density ($|B_{\textsc{ch}}|$) of the coronal hole as function of the area coverage of the assumed open field regions ($A_{\textsc{of}}$) as extracted using the threshold of $0.94 \times I_{\mathrm{\textsc{SD},med}}$. This is done for the whole sample and three latitude bins, namely CoM$_{\textsc{lat}} \leq 20^{\circ}$,  $20^{\circ} < $ CoM$_{\textsc{lat}} \leq 40^{\circ}$ and CoM$_{\textsc{lat}} > 40^{\circ}$. For the full sample we find a correlation of cc$_{\mathrm{sp}} = 0.59$, CI$_{90} = [0.52,0.66]$. The different latitude bins show a latitudinal dependence of the found relation with cc$_{\mathrm{sp}} = 0.74$, CI$_{90} = [0.66,0.81]$ in the lowest latitudinal bin which is associated with coronal holes near the solar equator. In the latitude bin covering latitudes from $20^{\circ}$ to $40^{\circ}$ a correlation of cc$_{\mathrm{sp}} = 0.50$, CI$_{90} = [0.37,0.61]$ was found. In the highest latitude bin ($>40^{\circ}$) the correlation vanishes (cc$_{\mathrm{sp}} = 0.01$, CI$_{90} = [-0.24,0.25]$). The correlations seem to follow an exponential curve and we use an exponential fit. For the whole dataset the fit is:
\begin{equation} \label{eq:fit}
    |B_{\textsc{ch}}| [\mathrm{G}] = (0.25 \pm 0.06) \times \mathrm{e}^{(0.032 \pm 0.003) \times A_{\textsc{of}} [\%]},
\end{equation}
and for the lowest latitude bin a slightly steeper fit with 
\begin{equation}
    |B_{\textsc{ch}}| [\mathrm{G}] = (0.13 \pm 0.04) \times \mathrm{e}^{(0.042 \pm 0.004) \times A_{\textsc{of}} [\%]},
\end{equation}
was found. With this relationship it is possible to estimate the open flux of a coronal hole by only using two spectral filters, the $193$\AA\ filter to define the coronal hole boundary and the $304$\AA\ filter to derive the open flux regions. Further, we estimate the uncertainty in the fit to be between $40\%$ for low open magnetic field areas and $60\%$ for higher ones (see Fig~\ref{fig:corrplot}, shaded area). The uncertainties were estimated by bootstrapping the fit over the sample with the uncertainties in B (from CATCH) in mind (see Section~\ref{subs:corr_methods}).   Due to the exponential nature, \textit{i.e.,} saturating behavior towards $100\%$ coverage, the uncertainty increases towards higher field strengths/higher coverage. 

\subsection{Far-Side Coronal Holes}
\begin{figure} 
 \centerline{\includegraphics[width=.9\textwidth,clip=]{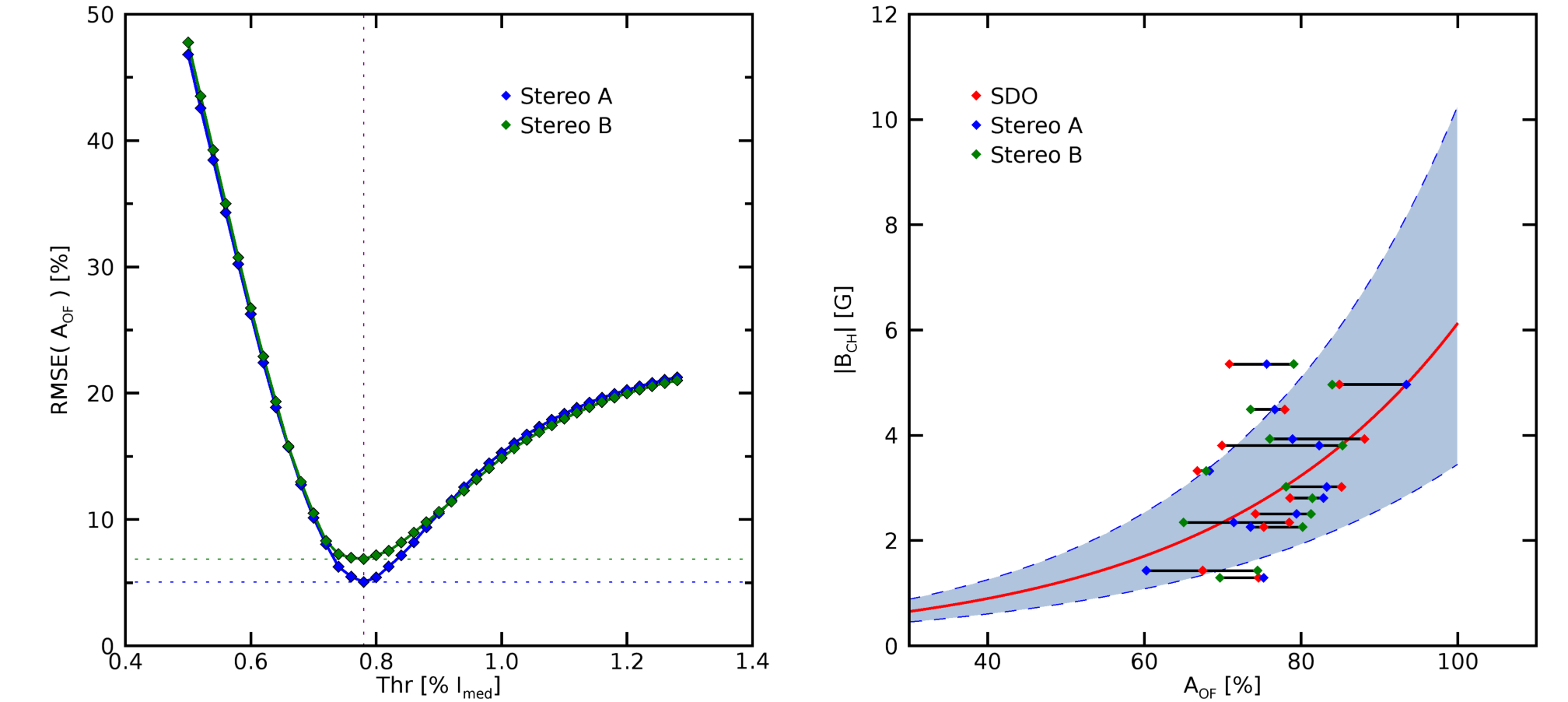}}
 \caption{Left: The RMSE between the percentage area coverage of STEREO and SDO is plotted as function of the $304$\AA\ threshold to determine the threshold values most closely correlated between SDO and STEREO. Green represents STEREO-B and blue shows STEREO-A. The purple dashed line the minimum at $0.78 \times I_{\mathrm{\textsc{SD},med}}$ for both STEREO-A (blue dashed line) and -B (green dashed line). Right: Coronal hole mean magnetic flux density ($|B_{\mathrm{\textsc{ch}}}|$)  as function of the percentage area coverage ($A_{\mathrm{\textsc{of}}}$) for the minimum threshold as marked in the left panel. The percentage area coverage is plotted for SDO (red), STEREO-A (blue) and STEREO-B (green), a single coronal hole is connected by the black lines. The shaded area and red line show the fit and uncertainties derived in Figure~\ref{fig:corrplot}, a) from the full coronal hole sample. }\label{fig:stereo_eval}
 \end{figure}

In Section~\ref{subs:ch+tr} we have established a relationship between the open magnetic field and the transition region intensity distribution in coronal holes. Using this relation we can estimate the open flux based on $304$\AA\ observations. Using the different perspectives from STEREO satellites allows to derive the magnetic flux of coronal holes from EUV observations at viewpoints not visible from Earth. Between $2012$ and $2017$, both STEREOs were separated by at least $110^{\circ}$ from Earth, effectively imaging the far-side of the Sun. As there is no remote sensing magnetic field instrument on either STEREO, this relation might be used to estimate the far-side open flux during that time. To be able to use this relation, a calibration from SDO to STEREO has to be established. For this, $13$ coronal holes of medium to large area were chosen, that were observed by both STEREOs and SDO, with a low longitudinal separation. The low separation angle between the spacecraft allows the approximation of a similar magnetic flux density. As the magnetic field evolution of a coronal hole is not dependent on the coronal holes area or lifetime evolution the statistical uncertainty may be even lower. Even though SDO and STEREO observe the same waveband ($304$\AA ), the instruments (e.g., filter, mirror, ccd, age) are not identical. The AIA/SDO EUV filters are built with a spectral Full Width Half Maximum (FWHM) of $\approx 1$nm \citep[][]{2014Boerner}, whereas the $304$\AA\ EUVI/STEREO filter have a FWHM of $3$nm \citep[][]{2008howard_SECCHI}. These differences make an inter-calibration necessary. For this, we use a simple approach, assuming that the relation found for SDO data is also valid for STEREO data. However the threshold for the detection of the supposedly open fields might not be the same. Thus, we calculated the root mean square error (RMSE) between the area coverage of STEREO and SDO in $304$\AA\ while varying the extraction threshold for STEREO, and keeping the SDO threshold fixed at $0.94 \times I_{\mathrm{\textsc{SD},med}}$ for SDO (Left panel of Figure~\ref{fig:stereo_eval}). From this we derive a minimum at  $0.78 \times I_{\mathrm{\textsc{SD},med}}$ for both STEREOs. We find that this threshold calibration yields similar percentage area ratios in $304$\AA\ ($A_{\textsc{of}}$) between the respective SDO and STEREO coronal holes (indicated by the black line connecting a red, black and blue point in the right panel of Figure~\ref{fig:stereo_eval}. For the thresholds $0.94$ and $0.78 \times I_{\mathrm{\textsc{SD},med}}$  for SDO and STEREO respectively, we find a mean error (ME) of ME$_{\textsc{sta}}= -0.7~\%A_{\textsc{of}}$ and ME$_{\textsc{stb}}= -0.2~\%A_{\textsc{of}}$, \textit{i.e,} the relative difference is $<1\%$. The RMSE was found to be RMSE$_{\textsc{sta}}= 5.0~\%A_{\textsc{of}}$ and RMSE$_{\textsc{stb}}= 6.8~\%A_{\textsc{of}}$, or a relative absolute difference of $6.6\%$ and $9.0\%$ respectively. To verify that the data from STEREO also follows the same relation as found from SDO data, we show on the right of Figure~\ref{fig:stereo_eval}, the signed mean magnetic flux density ($|B_{\textsc{ch}}|$) as function of the area coverage ($A_{\textsc{of}}$) for all three spacecraft. A single coronal hole is shown connected by a black line. We find that the extracted areas ratios for STEREO are matching the SDO area ratios very well and lie within the uncertainty range of the fit derived in Section~\ref{subs:ch+tr} (see Equation~\ref{eq:fit}). However we do note that, to independently fit the data obtained from STEREO observations or undoubtedly prove the relation for STEREO, the sample is too small and temporally too constrained. However, validity of the relation is suggested by the results. \\

The found results allow an estimation of the non-polar far-side open magnetic flux by first extraction coronal holes using \textit{e.g.,} the $195$\AA\ filter and then determining the percentage of the area below a threshold of $0.78 \times I_{\mathrm{\textsc{SD},med}}$ ($0.94 \times I_{\mathrm{\textsc{SD},med}}$ for SDO) in $304$\AA\ STEREO observations. This gives an estimate of the signed mean magnetic flux density including the uncertainties. From the coronal hole area and signed mean magnetic flux density the open flux for each coronal hole can be determined and subsequently the open far-side flux estimated.

\section{Discussion}\label{s:disc}
The difference between open flux observed remotely and measured in-situ is above a factor of $2$ \citep[][]{Linker_2017, 2017Lowder, 2019wallace}, which might partly be due to the uncertainty in estimating the far-side open flux from indirect methods like synoptic charts or flux-transport models. This new method to estimate the open flux from far-side remote sensing observations may allow closing the gap of the ``open flux problem" and help in validating new methods to estimate the far-side magnetic fields. With the methodology reported in this study, it is possible to derive the far-side open flux with an uncertainty between $40\%$ and $60\%$ (depending on the coverage) from the uncertainty in the magnetic flux density, which is close to the uncertainty in the open field of a single coronal holes observed at the near-side of $45\%$ as reported by \cite{2021Linker_ISSI}. However, the value obtained by \cite{2021Linker_ISSI} includes an estimate for the uncertainty in the coronal hole area ($\approx 26\%$), which was not considered in this study.  \\

In the photosphere we find a linear relation between the area covered by the magnetic elements and the mean magnetic flux density of the coronal hole, but in the chromosphere/transition region, we choose to use an exponential fit. In the photosphere the area of a magnetic element is determined entirely by its flux, or vice versa \citep[see][]{2019hofmeister}. In the transition region, we believe that the area covered by fields extending from those photospheric magnetic elements is determined primarily from the magnetic pressure balance between open and closed field. The interaction of loops from bipolar structures and open fields from unipolar funnels will determine the area ratio related to open fields \citep[see the theoretical concepts by][]{2005cranmer,2009wedemeyer}. It seems from Figure~\ref{fig:corrplot} that this relation can be described by an exponential function rather than a linear one as found in the photosphere.\\

We find that the relation between the area coverage and the magnetic field starts to decrease towards higher latitudes, which we strongly assume to be a projection effect rather than an actual physical effect. The AIA EUV filters cover emission from multiple atmospheric heights. Due to this, in combination with the view angle, the same structure might show different signatures depending on what direction and angle they are observed from. In the case of coronal holes we suppose that the area of bright structures depends on these, however in a non-linear way. We tried to correct for the latitudinal dependence (not shown in the paper) and found no correlation between extracted areas and latitudes of these bright structures. This leads to the assumption that also the orientation plays a role in how big the structure is when seen at different angles  \citep[][]{2007Feng,2008Aschwanden}. By understanding the projection effects, and finding a way to correct for it, it might be possible to significantly increase the correlations and thus lowering the uncertainties. Further, the increasing uncertainty in magnetic field observation due to the LoS angle \citep[][]{2013Petrie, 2015Petrie} and due to the zero point uncertainty \citep[][]{2021Li}, might contribute to the decrease of the correlation. Future data from missions, that will go out of the ecliptic (\textit{e.g.,} Solar Orbiter, Solaris), will help to further constrain the uncertainties.\\

Except of the data uncertainties, we also have to discuss the assumptions and simplifications we have made. Firstly we assumed that the field in the upper chromosphere/transition region is mostly filled by magnetic fields that eventually are considered open, or high loops rooted in magnetic elements. Although this was done in accordance with the models and ideas of the coronal structure as proposed by \cite{2005cranmer} and  \cite{2009wedemeyer}, it has not yet been conclusively shown if that is the case. Secondly, it was assumed that the relation between the coronal hole signed mean magnetic flux density and the area coverage is similar for SDO and STEREO. Although the 13 coronal holes are a subset of the full SDO sample and follow the correlation, it is not possible to derive the relation from this data alone for the STEREOs. This is due to the small sample of usable data, as there is only a short time window where both SDO and STEREO were observing at a low longitudinal separation. By increasing the dataset to include coronal holes at larger longitudinal separation angles of the spacecraft, we would also increase the uncertainty in the magnetic field values and area differences due to the coronal holes evolution to the point where a meaningful correlation can no longer be derived.  

\section{Summary and Conclusions}\label{s:sum}
In this study, we present a new method to estimate the far-side open flux from STEREO observations. We correlate magnetic field structures within coronal holes from the photosphere to the transition region and derive an estimation of the open flux from $304$\AA\ images. We calculate an inter-instrument calibration to allow the use of this method for STEREO coronal holes. Our study can be summarized as follows.

\begin{enumerate}
    \item We showed that the intensity distribution, and as such the correlation of dark regions in $304$\AA\ to the magnetic flux density in coronal holes, significantly differs to quiet Sun and plage regions. 
    
    \item We show that the area coverage of photospheric magnetic elements of coronal holes can be approximated in $304$\AA\ by the area proportion below a threshold of $78\%$ (STEREO) or $94\%$ (SDO) of the solar disk median intensity.

    \item We successfully calibrated the area ratios for SDO and STEREO (RMSE $< 10 \% A_{\textsc{of}}$; ME $< 1 \% A_{\textsc{of}}$). This allows the use of the found relation on STEREO data.   
   
    \item From the relation between the area coverage ($A_{\textsc{of}}$) in $304$\AA\ and signed mean magnetic flux density ($|B_{\textsc{ch}}|$), we derived that the open flux of a coronal hole can be approximated as:
   \begin{equation}
    |\Phi_{\textsc{ch}}|[\mathrm{Mx}] = (0.25 \pm 0.06) \times A_{\textsc{ch}} \times \mathrm{e}^{(0.032 \pm 0.003) \times A_{\textsc{of}} [\%]}.
\end{equation}

\end{enumerate}

This new method on how to estimate the far-side flux may help in deriving far-side fluxes from helioseismic data or tune flux transport models. It can help to improve the estimates of global open magnetic flux needed for space weather forecasting and as input for heliospheric MHD simulations \citep[\textit{e.g.,}][]{2018euhforia}. Finally, the additional information that can be obtained using this method could act as input for L5 pre-studies (\textit{e.g.,} in preparation of ESA's planned Lagrange L5 mission).

 
%
 \begin{acks}
 The SDO and STEREO image data are available by courtesy of NASA and the respective science teams. S.G.H. acknowledges the German Data Center for SDO (GDC-SDO) for providing SDO data. Inspired by the International Space Science Institute (ISSI, Bern) Team on “Magnetic open flux and solar wind structuring in interplanetary space” (2019-2020). 
 \end{acks}
%

\footnotesize\paragraph*{Disclosure of Potential Conflicts of Interest}
The authors declare that they have no conflicts of interest.

%
%
\bibliographystyle{spr-mp-sola}
%
%
%
%

\end{article} 
\end{document}